\documentclass{aa}
\usepackage{times,float}
\usepackage{epsfig}
\usepackage{graphics}
\usepackage{amssymb}
\usepackage{float}

\begin{document}

\title{GRB  020813:   polarization  in  the   case  of  a   smooth  optical
    decay\thanks{Based on  observations collected at  the European Southern
    Observatory,  Cerro Paranal  (Chile), ESO  programmes  69.D-0461(A) and
    69.D-0701(A).}}

\titlerunning{Optical polarimetric follow up of the GRB~020813 afterglow}
\author{J. Gorosabel\inst{1,2}
       \and E. Rol\inst{3}
       \and S. Covino\inst{4}
       \and A.J. Castro-Tirado\inst{1}
       \and J.M. Castro Cer\'on\inst{2}
       \and D. Lazzati\inst{5} 
       \and J. Hjorth\inst{6} 
       \and D. Malesani\inst{7}
       \and M. Della Valle\inst{8}
       \and S. di Serego Alighieri\inst{8}
       \and F. Fiore \inst{9}    
       \and A.S. Fruchter\inst{2}
       \and J.P.U. Fynbo\inst{10}
       \and G. Ghisellini\inst{4}
       \and P. Goldoni\inst{11}
       \and J. Greiner\inst{12}
       \and G.L. Israel\inst{9}
       \and L. Kaper\inst{3}
       \and N. Kawai\inst{13} 
       \and S. Klose\inst{14}
       \and C. Kouveliotou\inst{15}
       \and E. Le Floc'h\inst{11}
       \and N. Masetti\inst{16}
       \and F. Mirabel\inst{11}
       \and P. M\o ller\inst{17}
       \and S. Ortolani\inst{18}
       \and E. Palazzi\inst{16}
       \and E. Pian\inst{19}
       \and J. Rhoads\inst{2}
       \and G. Ricker\inst{20} 
       \and P. Saracco\inst{4}
       \and L. Stella\inst{9}
       \and G. Tagliaferri\inst{4}  
       \and N. Tanvir\inst{21}
       \and E. van den Heuvel\inst{3}
       \and M. Vietri\inst{22}
       \and P.M. Vreeswijk\inst{23}
       \and R.A.M.J. Wijers\inst{3}
       \and F.M. Zerbi\inst{4}}
\institute{ Instituto de Astrof\'{\i}sica de Andaluc\'{\i}a (IAA-CSIC),
 P.O. Box 03004, E--18080 Granada, Spain.
\and 
 Space  Telescope Science Institute, 3700  San Martin  Drive, Baltimore, MD
 21218-2463,   USA.
\and
 University of Amsterdam, Kruislaan 403, NL--1098 SJ Amsterdam, The
 Netherlands.
\and
 INAF, Osservatorio Astronomico di Brera, via E. Bianchi 46, I--23807 Merate (LC), Italy.
\and
 Institute of Astronomy, University of Cambridge, Madingley Road, UK--CB3 0HA Cambridge, UK.
\and
 Astronomical Observatory, University of Copenhagen,
 Juliane Maries Vej 30, DK--2100 Copenhagen \O, Denmark.
\and
 International  School for  Advanced Studies  (SISSA/ISAS), via  Beirut 2-4,
 I--34016 Trieste, Italy.
\and
 INAF, Osservatorio Astrofisico di Arcetri, Large E. Fermi 5, I--50125 Firenze,
Italy.
\and
 INAF,  Osservatorio   Astronomico  di   Roma,  via  Frascati   33,  I--00044
Monterporzio, Italy.
\and
 Department of Physics and Astronomy, University of \AA rhus,  Ny
 Munkegade, DK--8000 \AA rhus C, Denmark.
\and
 CEA/DSM/DAPNIA, L'Orme des Merisiers, Bat. 709, F--91191 Gif-sur-Yvette,
 France.
\and
 Max-Planck-Institut f\"ur extraterrestrische Physik, D--85741 Garching, Germany.
\and
 Department of  Physics, Faculty of  Science, Tokyo Institute  of Technology
 2-12-1 Oookayama, Meguro-ku, Tokyo 152-8551, Japan.
\and
 Th\"uringer Landessternwarte Tautenburg, D--07778 Tautenburg, Germany.
\and
 NASA MSFC, SD-50, Huntsville, AL 35812, USA.
\and
 Istituto di Astrofisica Spaziale e Fisica Cosmica, CNR, Via Gobetti 101, 
 I--40129 Bologna, Italy.
\and
 European Southern Observatory,
 Karl--Schwarzschild--Stra\ss e 2,
 D--85748 Garching, Germany.
\and
 Universit\`a di Padova, Dept. di Astronomia, Vicolo dell'Osservatorio 2,
 I--35122 Padova, Italy.
\and
 INAF, Osservatorio Astronomico di Trieste, Via G.B. Tiepolo 11, 
 I--34131 Trieste, Italy.
\and
Center   for  Space  Research,   Massachusetts  Institute   of  Technology,
Cambridge, MA 02139-4307, USA.
\and
 Department of   Physical Sciences, University  of Hertfordshire,
 College     Lane,    Hatfield,   Herts UK--AL10     9AB,  UK.
\and
 Scuola Normale Superiore, Piazza dei Cavalieri, I--56100 Pisa, Italy.
\and
European Southern Observatory, Casilla 19001, Santiago 19,
Chile
}
\offprints{ \hbox{J. Gorosabel, e-mail:{\tt jgu@iaa.es}}}

\date{Received / Accepted }

\abstract{We present the results  of a VLT polarimetric monitoring campaign
  of the  GRB~020813 optical  afterglow carried out  in three  nights, from
  0.88 to  4.20 days  after the  gamma-ray event.  The  mean values  of the
  degree  of linear polarization  ($P$) and  its position  angle ($\theta$)
  resulting from  our campaign are $<P>=1.18  \pm 0.10 \%$  and $<\theta> =
  148.7^{\circ}   \pm   2.3^{\circ}$,   after   correcting   for   Galactic
  interstellar polarization.  Our VLT  data set is most naturally described
  by  a  constant  degree   of  linear  polarization  and  position  angle,
  nonetheless a slow $\theta$ evolution cannot be entirely ruled out by our
  data.  The  VLT monitoring campaign  did not reveal either  a significant
  $\theta$  rotation respect to  the Keck  spectropolarimetric observations
  performed $\sim0.25$  days after the GRB (Barth  et al.  \cite{Bart03a}).
  However,  $<P>$ is a  factor of  two lower  than the  polarization degree
  measured from  Keck.  Additionally, the VLT polarization  data allowed us
  to construct an accurate  $V$-band light curve.  The $V$-band photometric
  data revealed a smooth light curve  with a break located between the last
  Keck and our first VLT polarimetric measurement, $0.33 < t_{\rm break, V}
  < 0.88$  days after the GRB.   The typical magnitude  fluctuations of the
  VLT $V$-band lightcurve are  $0.003$~mag, $0.010$~mag and $0.016$~mag for
  our  three  observing  nights,   respectively.   We  speculate  that  the
  stability of  $\theta$ might  be related to  the smoothness of  the light
  curve.   }

\maketitle

\section{Introduction}

GRB~020813  was  detected by  the  HETE-2  spacecraft  on Aug  13.11411  UT
(Villase\~nor et al.  \cite{Vill02}) as  a bright long (duration $>$ 125~s;
Hurley et  al. \cite{Hurl02})  gamma-ray burst (GRB)  and its  position was
rapidly disseminated among the  GRB community.  After the optical afterglow
(OA)  discovery (Fox, Blake  \& Price  \cite{Fox02}) it  was subject  of an
intensive  radio (Frail  \& Berger  \cite{Frai02}; Bremer  \& Castro-Tirado
\cite{Brem02}), X-ray  (Butler et al.   \cite{Butl03}), optical (Li et  al. 
\cite{Li03}; Laursen \& Stanek \cite{Laur03}; Urata et al.  \cite{Urat03}),
near-IR  (Covino  et  al.  \cite{Covi03a}), polarimetric  (Barth  et  al.
\cite{Bart02}; Covino  et al.   \cite{Covi02}) and spectroscopic  (Price et
al. \cite{Pric02}; Barth et al.  \cite{Bart03a}) follow-up.

Among  the thousands  of  GRBs detected  to  date, only  $\sim$50 of  those
accurately                                            localized\footnote{See
  http://www.mpe.mpg.de/$\sim$jcg/grbgen.html}  have   been  pinpointed  at
optical  wavelengths.  Positive  linear polarization  detections  have been
reported for 7 of them, typically at a level of $1-3\%$ (with the possible
exception of the $\sim10\%$ polarization reported for GRB~020405 by Bersier
et al.   \cite{Bers03}).  A review  on the polarization detections  to date
can  be  found  in  Covino   et  al.   (\cite{Covi03b})  and  Bj\"  ornsson
(\cite{Bjor02}).   Optical  afterglow emission  is  widely  accepted to  be
synchrotron  radiation,   the  result  of   the  interaction  of   the  GRB
relativistic wave with the circumburst  medium (fireball model; see Rees \&
M\'esz\'aros  \cite{Rees92} and  M\'esz\'aros \&  Rees  \cite{Mesz97}).  In
general, if  the fireball  configuration is not  symmetric, some  degree of
polarization is expected from the synchrotron radiation.
 
Several asymmetric scenarios able to  account for the $1-3\%$ polarization
typically  measured  in OAs  have  been  proposed.   Among them  are:  $i$)
causally disconnected magnetic  patches (Gruzinov \& Waxman \cite{Gruz99}),
$ii$) a  homogeneous collimated  fireball observed off-axis  (Ghisellini \&
Lazzati \cite{Ghis99};  Sari \cite{Sari99})  and $iii$) a  similar scenario
where the  jet is structured (the  collimated energy per  unit angle decays
smoothly with the angle from the jet axis; Rossi et al.  \cite{Ross02}).

In the context of $i$) $\theta$ and  $P$ are expected to change on the same
time  scale.    The  latter  two   scenarios  (~$ii$)  and   $iii$)~)  show
characteristic evolutions of  $\theta$ and $P$ when the  outflow slows down
and  the  Lorentz factor  decreases.   It  is  especially interesting  that
scenario  $ii)$  predicts  a  double-peaked  evolution  of  the  degree  of
polarization, with a polarization angle change of 90 degrees in between the
two  peaks, while  in  contrast scheme  $iii)$  predicts a  single peak  of
polarized  emission, with  a  constant polarization  angle.  If an  ordered
magnetic field exists  in the medium into which  the shock propagates, this
can  result  in a  $\theta$  roughly constant  in  time,  accompanied by  a
variable $P$ (Granot \& K\"onigl \cite{Gran03}).

Among the  7 positive OA polarization  detections there are  only two clear
cases  (GRB~021004 and  GRB~030329) where  a rotation  in  the polarization
angle  has  been  reported  (Rol  et  al.   \cite{Rol03};  Greiner  et  al.
\cite{Grei03}). In  both cases the  OA shows a  complex light curve.   So a
satisfactory description  of the  polarization evolution is  still pending.
In the present  paper we report an intensive polarimetric  follow up of the
GRB~020813 OA.  In a companion paper Lazzati et al.  (\cite{Lazz04}) report
a physical interpretation of the polarization data published in this study.

 \section{Observations}
 \label{Observations}
 
 The polarimetric  follow up observations  started $0.8795$ days  after the
 gamma-ray event.  The observations were performed in the $V$-band with the
 FOcal  Reducer/low dispersion  Spectrograph 1  (FORS1) at  the  Very Large
 Telescope (VLT), unit  3. The 2048 $\times$ 2048 FORS1  CCD yields a pixel
 scale of $0.2^{\prime \prime}$/pix, and  was used in the high gain 4-ports
 readout  mode. A  Wollaston prism  in  tandem with  a rotatable  half-wave
 retarder allowed us  to determine the Stokes $Q$  and $U$ parameters.  For
 each  retarder angle  $\phi/2$,  two orthogonal  simultaneous images  with
 polarization  angles $\phi$ and  $\phi +  90^{\circ}$ were  obtained.  Our
 polarimetric data are based on  four $\phi/2$ values (0.0, 22.5, 45.0, and
 67.5  degrees),  the observations  being  consecutive  executions of  four
 exposure  cycles.  The  log of  observations presented  in Table~\ref{log}
 contains 33 polarimetric cycles executed during three nights, amounting to
 $\sim13$ hr of VLT exposure time.

 \section{Analysis}
 
 The images were reduced following  standard procedures. First, for each of
 the  FORS1  images  the  chip  overscan  of the  four  readout  ports  was
 subtracted. This was  done because the four pedestal  levels vary slightly
 from one  image to another.   This procedure was  carried out for  all the
 science and calibration images (including the bias images).  Then a master
 bias frame was  constructed by median combination and  subtracted from the
 science and flat field images.  Finally the science images were divided by
 a master  normalized flat  field image.  The  master flat field  image was
 created combining  sky flat  field frames that  were acquired  without the
 Wollaston    prism    and    the    retarder   plate    in    the    light
 path\footnote{Following    the     recommendations    given    in    {\tt
     http://www.eso.org/instruments/fors1/pola.html}}.  This  procedure  was
 applied separately to the data taken each observing night.
 
 The photometry is based on circular aperture photometry, fixing the radius
 to the  OA Full Width  at Half Maximum  (FWHM).  The results  presented in
 this study remain qualitatively  unaltered for aperture radii ranging from
 0.5 to 3 times the  OA FWHM.  Verification of the photometry, calibration,
 and  the  reduction  procedure,  were  performed observing  each  night  a
 polarization  standard star; BD$-12^{\circ}5133$  (13-14/08/2002), Hiltner
 652 (14-15  and 16-17/08/2002).  The  photometry and the reduction  of the
 standard stars was identical to the one applied to the science images.
 
\begin{figure}[t]
\begin{center}
  {\includegraphics[width=\hsize]{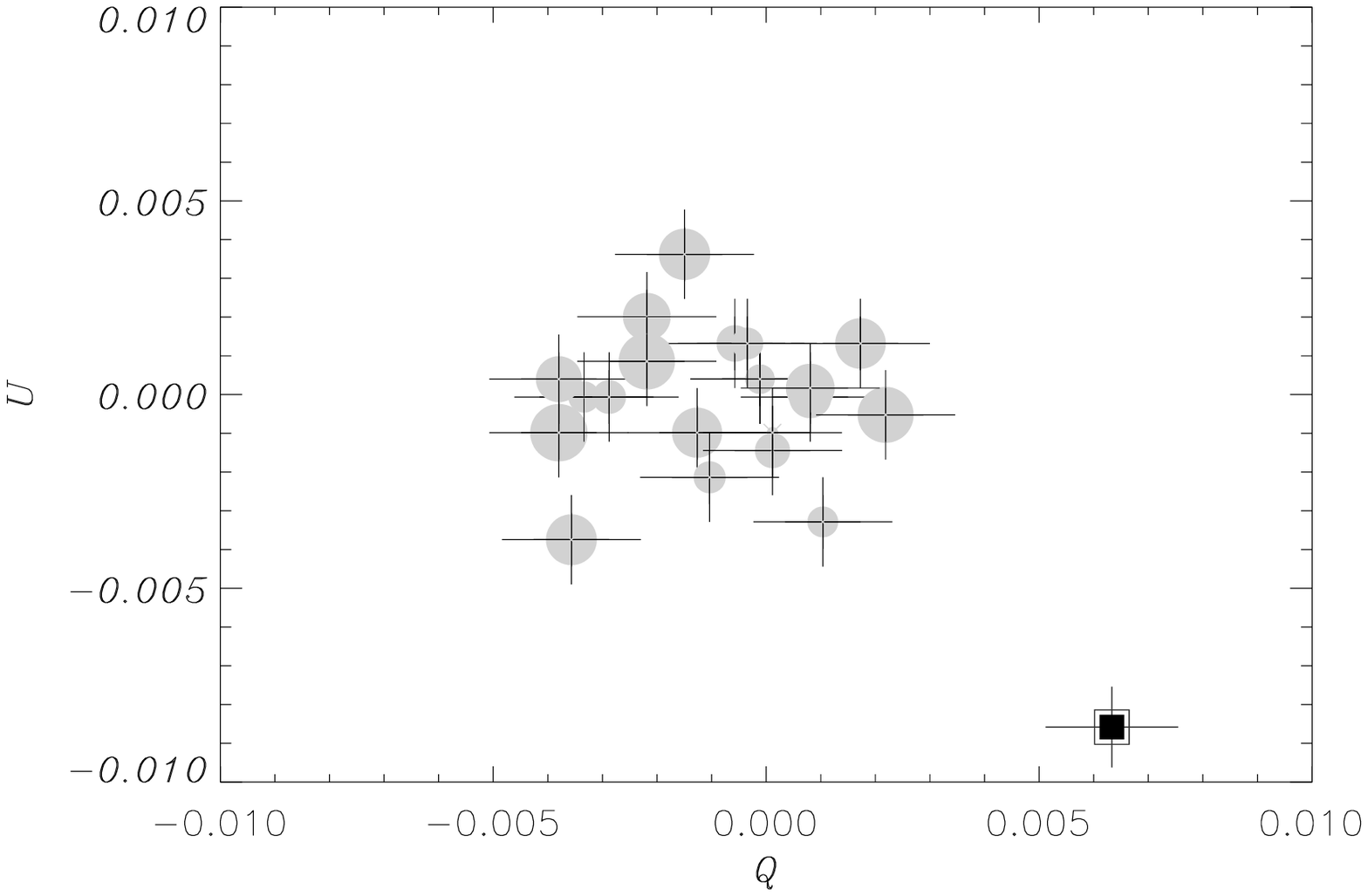}}
  {\includegraphics[width=\hsize]{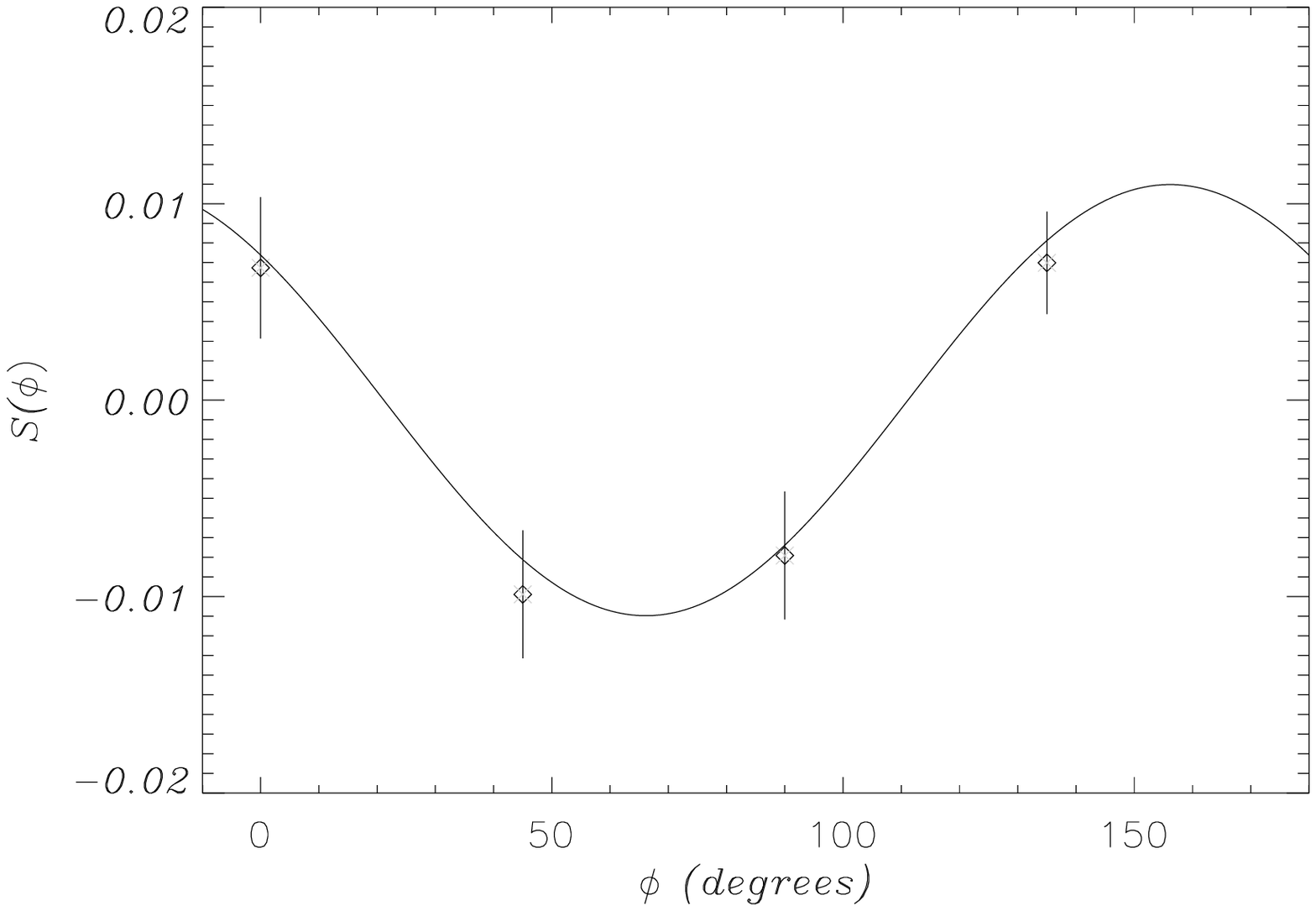}}
 \caption{\label{QU}  Both plots  correspond to  the  observations done 
   on Aug.   $13.99358 -  14.03042$ UT ($T_{exp}  = 4 \times  750$~s). {\bf
     Upper panel:} The diagram shows the Stokes ($Q$, $U$) parameters of 20
   field stars (circles)  and the OA (square), once  the ISM correction has
   been introduced.   The OA  is clearly polarized,  located away  from the
   diagram stellar  population.  The circle  sizes are proportional  to the
   distance  of each  star from  the  center of  the FORS1  image.  It  was
   noticed that the distance from the stars to the OA (placed in the center
   of the FORS1 CCD) is correlated to their polarization, presumably due to
   a residual  instrumental polarization effect.  Thus,  stars further than
   $1.7^{\prime}$ (not included  in the plot) from the  CCD center were not
   considered for the ISM correction.  {\bf Lower panel:} The plot displays
   the $S(\phi)= \frac{P(\%)}{100} \cos 2 (\theta - \phi)$ fit in agreement
   with the OA  position in the upper plot.   The fit yields $P=1.07\pm0.22
   \%$, $\theta= 154.3^{\circ}\pm5.9^{\circ}$, with $\chi^2/d.o.f=0.07$. }
\end{center}
\end{figure}

 The determination of the  parameters describing the linear polarization of
 ($P$ and $\theta$) is based  on the construction of the $S(\phi)$ function
 for the four $\phi/2$ retarder angles following the procedure described in
 Covino  et  al.   (\cite{Covi99}). Then  a  fit  of  the form  $S(\phi)  =
 \frac{P(\%)}{100}  \cos   2  (\theta  -   \phi)$  was  used   to  evaluate
 simultaneously  the  values  of  $P(\%)$  and  $\theta$.   An  independent
 verification  of  the  derived  $P$  and  $\theta$  values  was  performed
 evaluating the Stokes parameters ($Q$, $U$) for all the objects in the GRB
 field  based  on  Fourier  transformation arithmetics  (see  FORS1  manual
 expressions\footnote{FORS1+2 manual,  ref.  VLT-MAN-ESO-13100-1543, posted
   at  {\tt http://www.eso.org/instruments/fors/userman/}}).   Both methods
 yielded consistent results (see Fig.~\ref{QU}).

 The  Galactic interstellar  Medium  (ISM) reddening  in  the direction  of
 GRB~020813  is  not  negligible  ($E(B-V)=0.111$  mag;  Schlegel  et  al.  
 \cite{Sche98}), so it induces  (approximately) a systematic offset for all
 the  field objects  in the  Stokes ($Q$,  $U$) plane.   Thus, in  order to
 compensate the  effect of the ISM,  the weighted mass center  of the ($Q$,
 $U$) distribution was calculated for 20  stars in the field and shifted to
 the  Stokes plane  origin. We  realized that  stars located  close  to the
 border  of  the   FORS1  chip  show  a  residual   polarization  once  the
 interstellar polarization  effect has been  corrected (see upper  panel of
 Fig.~\ref{QU}).  Thus, stars located  further than $1.7^{\prime}$ from the
 image center were  not used for the Galactic  ISM polarization correction. 
 The   introduced   mean   offset   is   given   by   $\Delta   Q_{ISM}   =
 -6.22\pm0.59\times10^{-3}$        and       $\Delta        U_{ISM}       =
 3.95\pm0.80\times10^{-4}$.   These values  correspond  to $\theta_{ISM}  =
 178.2^{\circ}\pm2.6^{\circ}$  and  $P_{ISM}  =   0.62  \pm  0.06  \%$,  in
 agreement  with  the  ISM  correction   carried  out  by  Covino  et  al.  
 (\cite{Covi02}): $\theta_{ISM} = 178^{\circ}$ and $P_{ISM} = 0.59 \%$.  We
 have verified  that for the  applied ($\Delta Q_{ISM}$,  $\Delta U_{ISM}$)
 mean  offset the  field stars  remain, at  different epochs,  at  the same
 position  (within  errors)  on  the  Stokes  plane.   Empirical  relations
 (Serkowski,  Mathewson \& Ford  \cite{Serk75}) indicate  that $P_{ISM}(\%)
 \le 9 \times  E(B-V)$ mag, in agreement with the  derived $P_{ISM}$ in the
 direction of GRB~020813.
 
 Once  the  ISM  correction  is  introduced,  the value  of  $P$  has  been
 multiplied by  $\sqrt{1-(\sigma_P/P)^{2}}$, $\sigma_P$ being  the standard
 deviation of  $P$.  This  correction factor is  introduced because  $P$ is
 positive by definition, so averaged low signal-to-noise (S/N) polarization
 estimates (as  in our case) tend  to overestimate the actual  value of $P$
 (di   Serego   Alighieri    \cite{dise97}).    Although   FORS1   uses   a
 super-achromatic half-wave plate, it  introduces an offset in the $V$-band
 polarization angle of $1.80^{\circ}$, which  has to be subtracted from the
 inferred $\theta$ values.

\section{Results and discussion}
\label{Results}

\subsection{Determination of the break time}
\label{beuermann}

The accurate determination of the light curve break has an important impact
on the interpretation of the polarimetric evolution. However, the published
positions of the $V$-band break time  ($t_{\rm break, V}$) are not in close
agreement.   Thus,  Urata et  al.   (\cite{Urat03})  report $t_{\rm  break,
  V}=0.22\pm0.01$  days,  whereas  Covino  et  al.   (\cite{Covi03a})  give
$t_{\rm break, V}=0.50\pm0.27$ days.  In addition, Li et al.  (\cite{Li03})
determined an  $R$-band break time at $t_{\rm  break, R}=0.14\pm0.03$ based
on data  of the KAIT  telescope.  Again this  break time is  not consistent
with the $R$-band break  time ($t_{\rm break, R}=0.57\pm0.05$ days) derived
by Covino et al.  (\cite{Covi03a}).  Therefore, it is not clear whether the
spectropolarimetric  Keck observations carried  out $0.19-0.33$  days after
the  GRB (Barth  et al.   \cite{Bart03a})  were acquired  after, during  or
before the break.

Given the relevance  of a proper $t_{\rm break,  V}$ determination, we have
enhanced our  VLT sample with the published  GRB~020813 $V$-band magnitudes
to date  (Di Paola  et al.  \cite{DiPa02};  Covino et  al.  \cite{Covi03a};
Urata  et   al.   \cite{Urat03})  and   fitted  a  smoothly   broken  power
law\footnote{$F_{\nu}  =   (F_{1}^{-s}+F_{2}^{-s})^{-1/s}$,  with  $F_i=k_i
t^{\alpha_i}$  being the  pre  ($i=1$)  and post  ($i=2$)  break power  law
decays, and $s$ a non negative  number.  In this formalism $t_{\rm break} =
(\frac{k_1}{k_2})^{\frac{1}{\alpha_2-\alpha_1}}$}    (Beuermann    et   al.
\cite{Beue99}).  At  this point no host galaxy  contribution was considered
in the fit, repeating exactly the procedure previously carried out by other
authors  for  GRB~020813  (Covino  et  al.  \cite{Covi03a};  Urata  et  al.
\cite{Urat03}).  The compiled $V$-band data  points were shifted to our VLT
photometric  zero  point  (see  Sect.~\ref{lightcurve}).   The  fit  yields
$t_{\rm  break, V}=0.56\pm0.21$  days,  in agreement  with  the results  by
Covino et al.  (\cite{Covi03a}; $t_{\rm break, V}=0.50\pm0.27$ days).

We have checked the impact that the host galaxy contribution ($\sim25\%$ of
the total $V$-band flux of our last  VLT data points) might have on the fit
and therefore on the $t_{\rm break, V}$ determination. For this purpose two
additional late epoch  data points (where the host  is dominant) were added
to our data  set. First, the $BR$-band data points  acquired with the 1.54m
Danish Telescope (1.54dk) $\sim$58.5 days after the burst (Gorosabel et al.
\cite{Goro02}) were  interpolated in the  $V$-band yielding $V=24.2\pm0.2$.
Second we included in our  fits the HST/ACS $V$-band magnitude ($V=24.4 \pm
0.2$)  based on the  observations carried  out $70.03$  days after  the GRB
event\footnote{A  detailed  analysis  on   the  light  curve  and  the  HST
observations  (Based on  data  acquired under  Cycle  11 programme  \#9405,
P.I.:~Fruchter,~A.S.)  is beyond  the scope of the present  paper, and will
be published  elsewhere (Castro Cer\'on  et al. \cite{Cast04}).}.   In this
case   the  smoothly   broken   power  law   fit   yields  $t_{\rm   break,
V}=0.46\pm1.09$ days,  indicating the  lack of enough  $V$-band data  for a
simultaneous  fit in  the entire  parameter space  (given by  $k_1$, $k_2$,
$\alpha_1$, $\alpha_2$, $s$ and $V_{host}$).

\begin{figure}[h]
\begin{center}
  {\includegraphics[width=8cm]{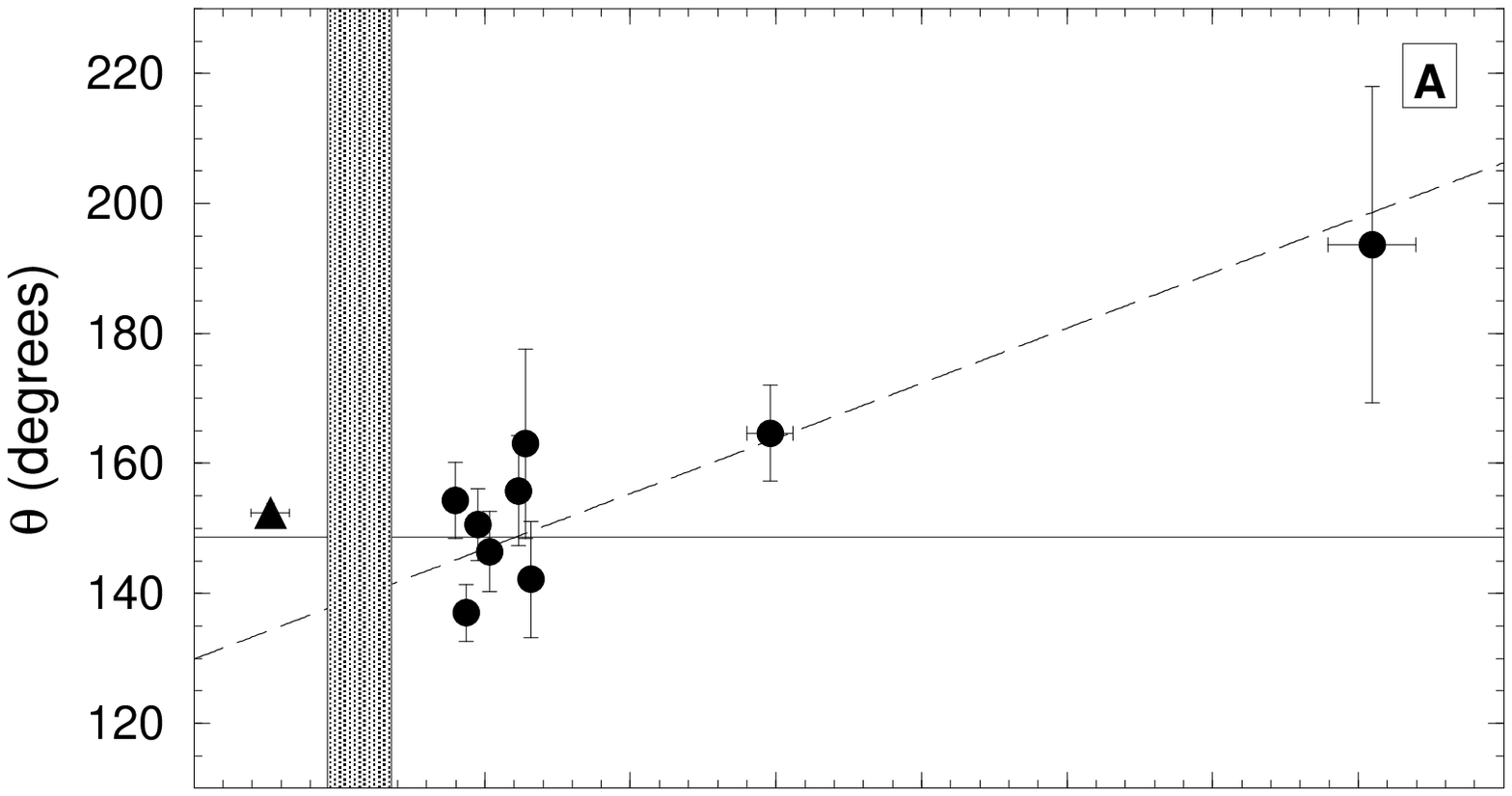}}
  {\includegraphics[width=8cm]{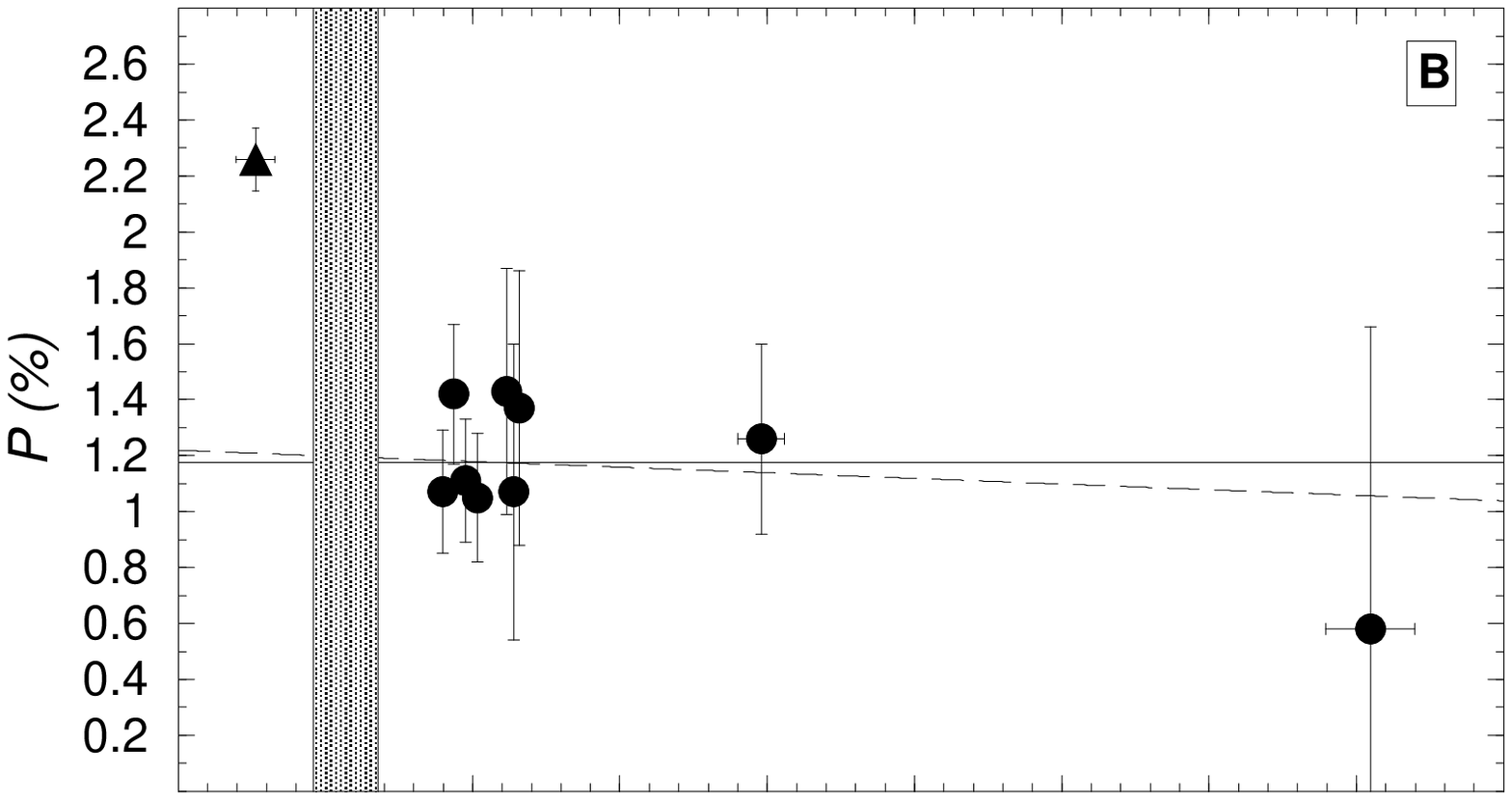}}
  {\includegraphics[width=8cm]{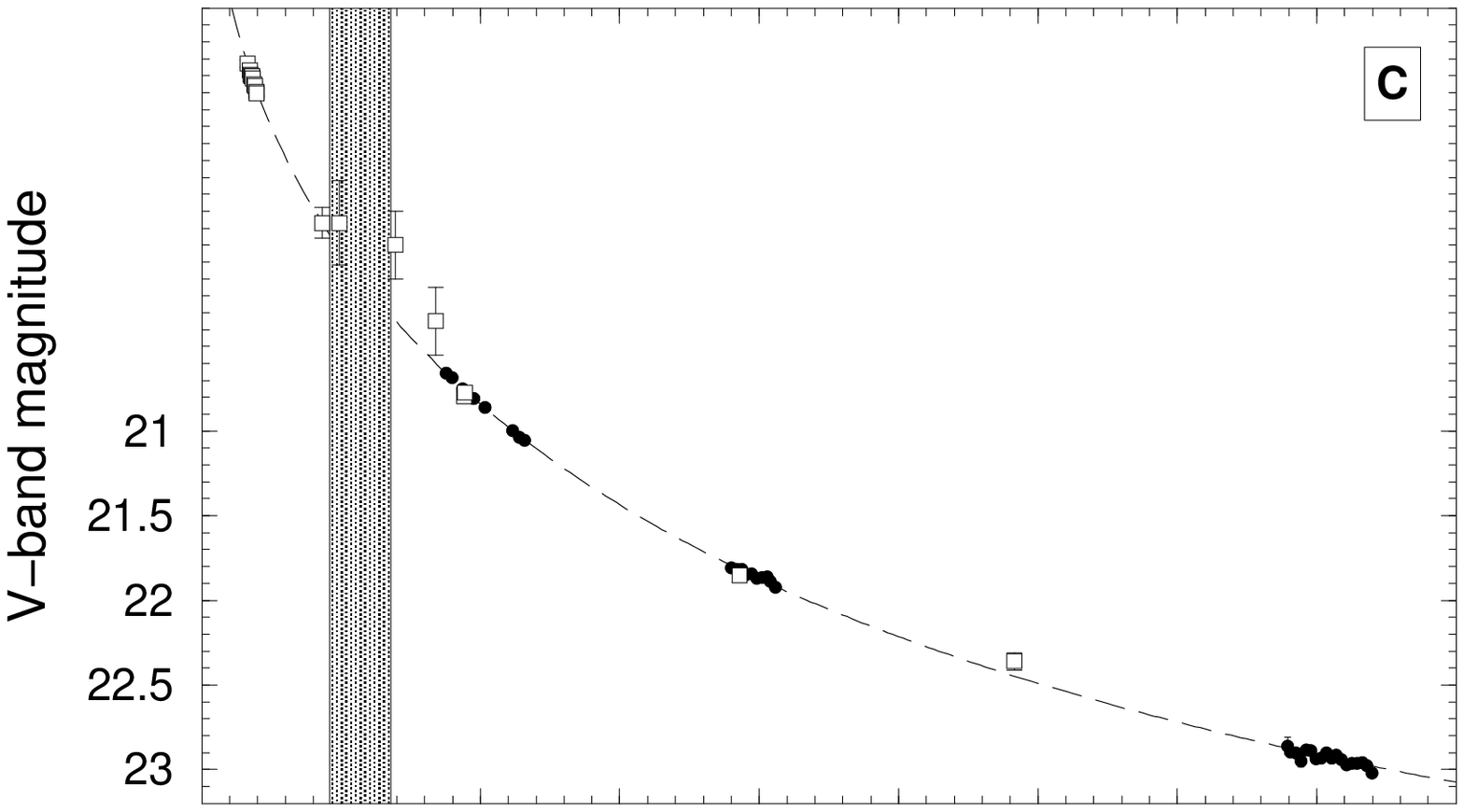}}
  {\includegraphics[width=8cm]{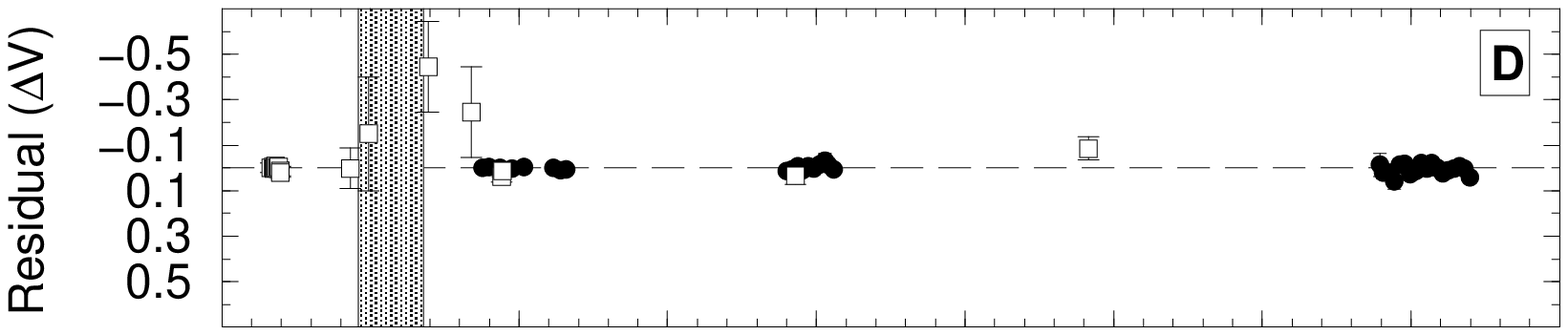}}
  {\includegraphics[width=8cm]{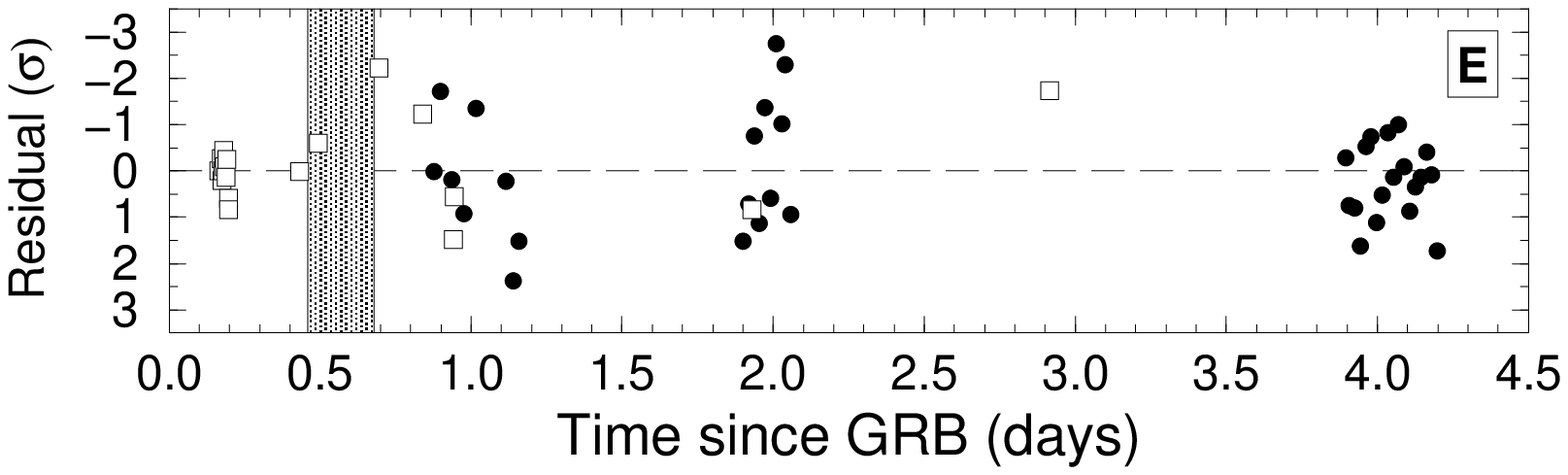}}
 \caption{\label{pola}  {\bf  (A) }  Evolution  of  the polarization  angle
   ($\theta$)  as   a  function  of   time.   The  circles   represent  our
   measurements and the  triangle the averaged Keck data  point.  The width
   of the error bars indicate  the duration of the observations.  The solid
   horizontal  line shows the  weighted VLT  mean and  the dashed  line the
   linear  fit to  the  VLT points.   {\bf(B)}  Linear polarization  degree
   ($P$), for the same points shown in the upper panel.  The horizontal and
   dashed  lines  represent the  same  as in  the  upper  plot.  {\bf  (C)}
   $V$-band  light  curve, based  on  the  36  VLT individual  data  points
   displayed  in Table~\ref{log} and  on the  measurements reported  in the
   literature.  The  dashed line  shows the smoothly  broken power  law fit
   when  $s=1$.  {\bf  (D)} $V$-band  magnitude residuals  when  the fitted
   light curve is  subtracted.  {\bf (E}) The residuals  expressed in units
   of standard  deviations.  {\bf General:} The vertical  shaded area shows
   the  break  time  $1\sigma$  uncertainty  region  ($s=1,  t_{\rm  break,
   V}=0.57\pm0.11$  days). The  filled circles  represent  the  VLT
   measurements  and the  empty  squares  the data  points  taken from  the
   literature.}
\end{center}
\end{figure}

In order to reduce the number of  parameters in the fit $s=1$ was fixed, as
previously carried out for other  OAs (Israel et al.  \cite{Isra99}; Stanek
et al.  \cite{Stan99}).  Then, leaving  $V_{host}$ as a free parameter, the
fit  is improved ($\chi^2/d.o.f=1.22$),  yielding; $\alpha_1=-0.55\pm0.09$,
$\alpha_2=-1.75\pm0.08$,   $t_{\rm   break,   V}=0.57\pm0.11$   days,   and
$V_{host}=24.61\pm0.17$  (see  Fig.~\ref{pola}   panel  C).   The  inferred
$t_{\rm break, V}$ value again supports the $V$-band break time reported by
Covino et al.  (\cite{Covi03a}), reducing its error by a factor of $\sim3$.
The  shaded  vertical band  of  Fig.~\ref{pola}  indicates  the $1  \sigma$
uncertain  region of  $t_{\rm  break,  V}$ when  $s=1$.   The exercise  was
repeated  for $s>1$  values, which  implied higher  $t_{\rm break,  V}$ and
$\chi^2/d.o.f$  values, hence  degrading the  fits.   In any  case for  any
reasonable value of $s$  ($1 < s < 30$) the inferred  break time is $t_{\rm
  break, V}  < 0.8795$ days,  epoch when our VLT  polarimetric observations
started.  For $s>30$  values the fits progressively disagree  with the data
($\chi^2/d.o.f > 2.0$).

Thus we conclude  that, independently of the host  galaxy magnitude and the
$s$ value, the Keck observations  (Barth et al.  \cite{Bart03a}) were taken
before  $t_{\rm break, V}$.   Therefore, the  $V$-band break  time occurred
between the last Keck and the  first VLT data points, $0.33 < t_{\rm break,
  V} < 0.88$  days after the GRB.  This result is  in disagreement with the
earlier  $t_{\rm  break,  V}$  determinations  inferred by  Urata  et  al.  
(\cite{Urat03}) and also with the $R$-band break epoch derived by Li et al.
(\cite{Li03}).

\subsection{The smooth $V$-band light curve}
\label{lightcurve}
In  order to  minimize the  relative magnitude  offset present  between our
$V$-band data points, we have fixed the photometric zero point with respect
to a bright unsaturated star  detected at high S/N ratio level (statistical
photometric errors  $< 0.001$).  The zero  point has been tied  to the star
located  at  $\alpha_{J2000}=  19^{h} 46^{m}  40.88^{s}$,  $\delta_{J2000}=
-19^{\circ}  35^{\prime}  15.9^{\prime  \prime}$  which has  $V=18.535  \pm
0.008$ (Henden \cite{Hend02}).  Thus, our light curve can be easily shifted
to  other  $V$-band  zero  points.   The  magnitude  errors,  displayed  in
Table~\ref{log}, include the statistical photometric error of the reference
star.

We have analyzed  the fluctuations of the $V$-band  light curve subtracting
from  the VLT  data  points the  function  fitted in  Sect.~\ref{beuermann}
($s=1$).  The residuals displayed in  Fig.~\ref{pola} panel D show that the
VLT points  (filled circles) follow  an extremely smooth light  curve, even
considering the small  photometric errors of the first  night.  The typical
magnitude deviations  of the VLT data  points from the  fitted function for
our three  observing nights  are $0.003$~mag, $0.010$~mag  and $0.016$~mag,
respectively.

 For  completeness we  have overplotted  in Fig.~\ref{pola}  (see  the open
    squares of  panels C,  D and E)  the $V$-band measurements  reported by
    other  authors   (Di  Paola  et  al.   \cite{DiPa02};   Covino  et  al.
    \cite{Covi03a}; Urata  et al.  \cite{Urat03}).  These  points are based
    on  different  telescopes/detectors so  very  likely  are  not free  of
    relative colour  term offsets.  Moreover,  we can not assure  that they
    were  obtained  using  one  consistent  photometric  technique,  so  in
    principle they are  expected to be more scattered  than our homogeneous
    VLT data set.

Considering  all the photometric  measurements (VLT  + other  authors), the
maximum residual  corresponds to a  $2.74 \sigma$ deviation,  associated to
the VLT  image taken on  $15.11512-15.13116$ UT (see  Fig.~\ref{pola} panel
E).  These results  suggest that the smooth light  curve measured $3.9-4.9$
hours after  the burst (see  Laursen \& Stanek \cite{Laur03})  continued at
least until 4.2 days after the GRB event.

\subsection{Polarization monitoring}

Barth et al.  (\cite{Bart03a}) reported  a potential rapid evolution of $P$
between   $0.19-0.33$  days  after   the  GRB   in  the   $4500-5500$  \AA~
spectropolarimetric wavelength interval.   However, this evolution vanishes
in  the  $5800-6800$  \AA~  bin,  so  these  authors  caution  against  any
overinterpretation on a possible  wavelength dependent $P$ variation (Barth
\cite{Bart03b}).  The FORS1 $V$-band  filter transmission curve is centered
at 5540.0 $\pm$ 557.5 \AA, just in the $5500-5800$ \AA~ gap present between
the two  spectropolarimetric bins.   Thus, in order  to extract  a $V$-band
filtered polarimetric data point from the spectropolarimetric data, we have
averaged the $P, \theta$ values  by Barth et al.  (\cite{Bart03a}) reported
for the two  wavelength ranges.  The averaged $P,  \theta$ values have been
derived using as a weight the area of the $V$-band filter response curve in
the  two  spectropolarimetric  bins\footnote{Assuming  a  $S(\lambda)  \sim
  \exp^{  -\ln(2)  (\frac{\lambda-5540}{557.5})^2}$  Gaussian profile,  the
  weights  were proportional  to $\int^{5500  \AA}_{4500 \AA}  S(\lambda) d
  \lambda$   and  $\int^{6800  \AA}_{5800   \AA}  S(\lambda)   d  \lambda$,
  respectively}.  Furthermore given the  unclear time evolution of $P$, the
three  epoch  data points  by  Barth  et  al.  (\cite{Bart03a})  have  been
averaged in time.  Finally, for consistency we corrected the averaged ($P$,
$\theta$) values for the slightly different ISM correction applied by Barth
et    al.     (\cite{Bart03a})   in    comparison    to    our   VLT    ISM
correction\footnote{Barth   et  al.    (\cite{Bart03a})   applied  an   ISM
  correction   given   by   $P^{Keck}_{ISM}   =  0.67   \pm   0.01\%$   and
  $\theta^{Keck}_{ISM}=167.0^{\circ}\pm0.3^{\circ}$,      in     reasonable
  agreement    with   $P^{VLT}_{ISM}    =   0.62    \pm   0.06    \%$   and
  $\theta^{VLT}_{ISM}=178.2^{\circ}\pm2.6^{\circ}$  used   in  the  present
  work.}.   We obtain  mean values  of $<P_{Barth}>  = 2.26  \pm  0.11 \%$,
$<\theta_{Barth}>   =   152.4^{\circ}  \pm   1.4^{\circ}$   for  the   Keck
observations.

The results  of our monitoring  campaign are displayed in  Table~\ref{log}. 
Given  the insufficient  OA S/N  ratio in  the individual  frames  taken on
14-15/08/2002  and 16-17/08/2002, the  data taken  within these  two nights
have been  co-added resulting in  two single polarimetric data  points.  We
have reported  in panels A and  B of Fig.~\ref{pola}  our nine polarimetric
measurements  (filled  circles)  along   with  the  mean  Keck  data  point
(triangle).  The VLT polarimetric  points might suggest a slowly increasing
trend  of $\theta$  with time  ($t$).  A  linear regression  to $\theta(t)$
provides a  satisfactory fit ($\chi^2/d.o.f=1.34$) consistent  with a slope
of  $17.0 \pm 5.6$  degree/day.  Thus,  the zero  slope constant  line (see
horizontal   solid  line  of   Fig.~\ref{pola}  panel   A)  departs   at  a
$\sim3\sigma$  level  from the  linear  fit  (dashed  line).  However,  two
arguments point against the suggestive smooth $\theta$ variation.

First, our  last inferred value of $\theta$  (Aug.  $17.01398-17.32148$ UT)
is based  on a  measurement which is  consistent with no  polarization (see
Fig.~\ref{pola}  panel B).   Second, the  fitted linear  $\theta$ evolution
predicts  for the Keck  mean observing  epoch (triangle  of Fig.~\ref{pola}
panel A)  $\theta=134.3^{\circ} \pm 6.8^{\circ}$,  underestimating the mean
Keck polarization angle ($<\theta_{Barth}>= 152.4^{\circ} \pm 1.4^{\circ}$)
at  a $2.6  \sigma$  level.   In contrast,  a  constant polarization  angle
scenario would naturally  match with the three $\theta$  values reported by
Barth et  al.  (\cite{Bart03a}).  The polarization degree  measured for the
nine VLT points  is fully consistent with no evolution,  since a linear fit
to  $P(t)$  yields  a  slope  of  $-0.04 \pm  0.25$  (see  dashed  line  of
Fig.~\ref{pola}  panel  B).  Thus,  our  VLT data  are  more  difficult  to
accommodate  in contexts  where  $P(t)$ and/or  $\theta(t)$ suffer  violent
rapid fluctuations.  However,  we cannot exclude that between  the epoch of
our  data points  (especially  in  the gap  between  Aug.~$15.18026$ UT  and
Aug.~$17.01398$ UT) there might  have occurred episodes of rapid variations
in $P(t)$ and/or $\theta(t)$.

Thus, within the  accuracy, sampling, and coverage of our  VLT data, we can
claim  that  there  is no  rotation  of  90  degrees  with respect  to  the
polarization  angle reported  before the  light curve  break (Barth  et al.
\cite{Bart02,Bart03a}) as  predicted in the context  of several theoretical
models  (i.e.   the nonspreading  homogeneous  jet,  Ghisellini \&  Lazzati
\cite{Ghis99}).

Qualitatively several models might be  accommodated in the framework of the
Keck plus VLT  data set: $i$) the lack of  polarimetric data $\sim$0.5 days
after  the burst  does not  allow  us to  exclude a  potential $P(t)$  peak
coincident  with the  light curve  break.   Hence, the  single $P(t)$  peak
predicted by Rossi et al.  (\cite{Ross02}) might still be consistent with a
$P(t)$ maximum placed  between the last Keck and our  first VLT data point.
$ii$) If the jet axis was close  to the observer line of sight, then also a
laterally spreading homogeneous jet might  show a single $P(t)$ peak placed
at  the light  curve break,  that  might agree  with the  joint data  (Sari
\cite{Sari99}).  $iii$) A third model potentially compatible with the whole
data set might be the one  based on a dominant large-scale ordered magnetic
field, which would  exist in the ISM where the  shock propagates (Granot \&
K\"onigl \cite{Gran03}).  In this case  a variable $P(t)$ is accompanied by
a roughly constant $\theta(t)$, as seen in GRB~020813.


\section{Conclusion}

To  date,  GRB  polarimetric  light  curves  have  been  sparsely  sampled,
especially  after the  break.  Several  OAs  have shown  small scale  rapid
optical fluctuations  with respect  to the canonical  power law  decay, but
very few have been monitored polarimetrically.

The only two afterglows which have shown a clear $\theta$ evolution to date
(GRB~021004,  Rol   et  al.   \cite{Rol03};  GRB~030329,   Greiner  et  al.
\cite{Grei03}), are  accompanied by structured light  curves, showing clear
deviations from smooth optical decays (Holland et al.  \cite{Holl03}; Guziy
et  al.  \cite{Guzi04}; Lipkin  et al.   \cite{Lipk03}).  In  contrast, the
GRB~020813  optical afterglow  shows  a  highly smooth  light  curve and  a
constant polarization angle,  as it has been reported  in the present work.
It is  suggestive to  speculate whether the  smoothness of the  light curve
might be correlated  with the stability of $\theta  (t)$.  This speculation
is  only based  on  the three  mentioned  cases, so  a further  statistical
verification  would  require an  intensive  polarization  monitoring for  a
significant sample of OAs.

We refer  to the companion  paper by Lazzati  et al. (\cite{Lazz04})  for a
detailed theoretical discussion and  interpretation of the data reported in
the present work.

 \begin{table*}[b]
 \begin{center}
   \caption{\label{log} Log of the observations carried out with the VLT(+FORS1).
     The  table is  divided in  three sub  tables, one  for  each observing
     night.  At the beginning of  each night the magnitude corresponding to
     a short  acquisition image ($T_{exp}$ = 20--150~s)  is displayed.  The
     polarization images  are summarized in  blocks of 4 images  cycles ($4
     \times T_{exp}$) with the FORS1  half wave plate rotator at 0.0, 22.5,
     45.0 and 67.5 degrees.}
 \begin{tabular}{lcccccc}
 \hline
 Date UT  &$T_{exp}$ & Filter & Seeing &  Magnitude & $P$   &$\theta$\\
 Aug. 2002&  (s)     &        & (``)   &            & (\%)  & (degrees)  \\
 \hline
  13.99037 - 13.99071 &$30$          &$V$&$0.53$& $20.6560 \pm 0.0190$&$\#$ &$\#$\\
  13.99358 - 14.03042 &$4 \times 750$&$V$&$0.64$& $20.6855 \pm 0.0022$&$1.07\pm0.22$&$154.3\pm5.9$\\
  14.03233 - 14.06918 &$4 \times 750$&$V$&$0.72$& $20.7485 \pm 0.0022$&$1.42\pm0.25$&$137.0\pm4.4$\\
  14.07107 - 14.10791 &$4 \times 750$&$V$&$0.68$& $20.8070 \pm 0.0022$&$1.11\pm0.22$&$150.5\pm5.5$\\
  14.11190 - 14.14869 &$4 \times 750$&$V$&$0.70$& $20.8600 \pm 0.0022$&$1.05\pm0.23$&$146.4\pm6.2$\\
  14.22276 - 14.23876 &$4 \times 300$&$V$&$0.84$& $20.9990 \pm 0.0036$&$1.43\pm0.44$&$155.8\pm8.5$\\
  14.24506 - 14.26128 &$4 \times 300$&$V$&$0.84$& $21.0355 \pm 0.0036$&$1.07\pm0.53$&$163.0\pm14.6$\\
  14.26340 - 14.27943 &$4 \times 300$&$V$&$0.74$& $21.0555 \pm 0.0036$&$1.37\pm0.49$&$142.1\pm8.9$\\
  \hline                                        
  15.00746 - 15.02343 &$4 \times 300$&$V$&$0.65$& $21.8081 \pm 0.0085$&$1.26\pm0.34^{\dagger}$&$164.7\pm7.4^{\dagger}$\\
  15.02526 - 15.04125 &$4 \times 300$&$V$&$0.73$& $21.8150 \pm 0.0078$& -  & - \\
  15.04323 - 15.05923 &$4 \times 300$&$V$&$0.74$& $21.8175 \pm 0.0081$& -  & - \\
  15.06114 - 15.07715 &$4 \times 300$&$V$&$0.74$& $21.8462 \pm 0.0074$& -  & - \\
  15.07915 - 15.09516 &$4 \times 300$&$V$&$0.68$& $21.8436 \pm 0.0060$& -  & - \\
  15.09722 - 15.11326 &$4 \times 300$&$V$&$0.59$& $21.8696 \pm 0.0064$& -  & - \\
  15.11512 - 15.13116 &$4 \times 300$&$V$&$0.60$& $21.8650 \pm 0.0053$& -  & - \\
  15.14428 - 15.14451 &$20$          &$V$&$0.55$& $21.8630 \pm 0.0322$&$\#$&$\#$\\
  15.14650 - 15.16247 &$4 \times 300$&$V$&$0.62$& $21.8870 \pm 0.0071$& -  & - \\
  15.16428 - 15.18026 &$4 \times 300$&$V$&$0.67$& $21.9229 \pm 0.0067$& -  & - \\
  \hline                                        
  16.98821 - 16.98995 &$150$         &$R$&$0.52$& $22.3220 \pm 0.0605$&$\#$ &$\#$\\  
  17.01049 - 17.01223 &$150$         &$V$&$0.60$& $22.8600 \pm 0.0502$&$\#$     &$\#$\\
  17.01398 - 17.03003 &$4 \times 300$&$V$&$0.66$& $22.8986 \pm 0.0271$&$0.58\pm1.08^{\star}$&$13.7\pm24.4^{\star}$\\
  17.03193 - 17.04800 &$4 \times 300$&$V$&$0.69$& $22.9036 \pm 0.0237$& -  & - \\
  17.05027 - 17.06634 &$4 \times 300$&$V$&$0.84$& $22.9500 \pm 0.0361$& -  & - \\
  17.06817 - 17.08429 &$4 \times 300$&$V$&$0.87$& $22.8842 \pm 0.0255$& -  & - \\
  17.08614 - 17.10224 &$4 \times 300$&$V$&$0.82$& $22.8863 \pm 0.0240$& -  & - \\
  17.10488 - 17.12100 &$4 \times 300$&$V$&$0.83$& $22.9385 \pm 0.0248$& -  & - \\
  17.12294 - 17.13906 &$4 \times 300$&$V$&$0.84$& $22.9307 \pm 0.0260$& -  & - \\
  17.14094 - 17.15707 &$4 \times 300$&$V$&$0.73$& $22.9031 \pm 0.0245$& -  & - \\
  17.15920 - 17.17533 &$4 \times 300$&$V$&$0.88$& $22.9339 \pm 0.0290$& -  & - \\
  17.17720 - 17.19334 &$4 \times 300$&$V$&$0.80$& $22.9127 \pm 0.0234$& -  & - \\
  17.19529 - 17.21146 &$4 \times 300$&$V$&$0.78$& $22.9402 \pm 0.0243$& -  & - \\
  17.21335 - 17.22953 &$4 \times 300$&$V$&$0.81$& $22.9716 \pm 0.0266$& -  & - \\
  17.23241 - 17.24859 &$4 \times 300$&$V$&$0.70$& $22.9640 \pm 0.0261$& -  & - \\ 
  17.25048 - 17.26666 &$4 \times 300$&$V$&$0.78$& $22.9646 \pm 0.0234$& -  & - \\
  17.26900 - 17.28521 &$4 \times 300$&$V$&$0.88$& $22.9595 \pm 0.0194$& -  & - \\
  17.28713 - 17.30334 &$4 \times 300$&$V$&$1.02$& $22.9752 \pm 0.0212$& -  & - \\
  17.30525 - 17.32148 &$4 \times 300$&$V$&$1.00$& $23.0218 \pm 0.0244$& -  & - \\
  \hline
  \multicolumn{7}{l}{$\#$      Not applicable.}\\
  \multicolumn{7}{l}{$\dagger$ Value obtained when co-adding all the polarimetric images taken on Aug. 15.00746-15.18026 UT.}\\
  \multicolumn{7}{l}{$\star$   Value obtained when co-adding all the polarimetric images taken on Aug. 17.01398-17.32148 UT.}\\
 \hline
 \end{tabular}
 \end{center}
 \end{table*}

\section*{Acknowledgments}

We  thank Aaron  Barth for  valuable information  on the  reduction  of the
GRB~020813 spectropolarimetric data.  The authors acknowledge benefits from
collaboration within  the Research Training Network  ``Gamma-Ray Bursts: An
Enigma and  a Tool'', funded  by the EU under  contract HPRN-CT-2002-00294.
We acknowledge our anonymous referee for his/her useful comments.

\end{document}